\documentclass[aps,prd,preprintnumbers,groupedaddress,nofootinbib,amssymb,eqsecnum,notitlepage]{revtex4}
\usepackage{graphicx}
\usepackage{bm}
\usepackage{amsmath}
\usepackage{color}
\usepackage{amsfonts}
\usepackage{here}
\usepackage{graphicx}
\usepackage{amsmath,amsthm,amssymb}
\usepackage{bm}
\allowdisplaybreaks[1]

\usepackage{amsfonts}
\usepackage{dcolumn}
\usepackage{hyperref}

\begin{document}
\newcommand{\newc}{\newcommand}

\newcommand{\ben}{\begin{eqnarray}}
\newcommand{\een}{\end{eqnarray}}
\newc{\be}{\begin{equation}}
\newc{\ee}{\end{equation}}
\newc{\ba}{\begin{eqnarray}}
\newc{\ea}{\end{eqnarray}}
\newc{\bea}{\begin{eqnarray*}}
\newc{\eea}{\end{eqnarray*}}
\newc{\da}{\delta{A}}
\newc{\D}{\partial}
\newc{\ie}{{\it i.e.} }
\newc{\eg}{{\it e.g.} }
\newc{\etc}{{\it etc.} }
\newc{\etal}{{\it et al.}}
\newcommand{\nn}{\nonumber}
\newc{\ra}{\rightarrow}
\newc{\lra}{\leftrightarrow}
\newc{\lsim}{\buildrel{<}\over{\sim}}
\newc{\gsim}{\buildrel{>}\over{\sim}}
\newc{\aP}{\alpha_{\rm P}}
\newc{\Mpl}{M_{\rm pl}}
\newc{\tb}{\tilde{\beta}}
\newc{\bb}{\bar{\beta}}
\newcommand{\rk}[1]{\textcolor{blue}{#1}}

\title{Odd-parity stability of hairy black holes 
in $U(1)$ gauge-invariant \\
scalar-vector-tensor theories}

\author{
Lavinia Heisenberg$^{1}$, 
Ryotaro Kase$^{2}$, and 
Shinji Tsujikawa$^{2}$}

\affiliation{
$^1$Institute for Theoretical Studies, ETH Zurich, Clausiusstrasse 47, 8092 Zurich, Switzerland\\
$^2$Department of Physics, Faculty of Science, Tokyo University of Science, 1-3, Kagurazaka,
Shinjuku-ku, Tokyo 162-8601, Japan}

\date{\today}

\begin{abstract}

In scalar-vector-tensor theories with $U(1)$ gauge 
invariance, it was recently shown that there exists a new type of 
hairy black hole (BH) solutions induced by a cubic-order 
scalar-vector interaction. In this paper, we derive conditions 
for the absence of ghosts and Laplacian instabilities against 
odd-parity perturbations on a static and spherically symmetric 
background for most general $U(1)$ gauge-invariant 
scalar-vector-tensor theories with second-order equations 
of motion. We apply those conditions to hairy BH solutions arising 
from the cubic-order coupling and show that the odd-parity stability 
in the gravity sector is always ensured 
outside the event horizon with the speed of gravity 
equivalent to that of light. We also study the case in which 
quartic-order interactions are present in addition to the 
cubic coupling and obtain conditions under which 
black holes are stable against odd-parity perturbations.

\end{abstract}


\maketitle

\section{Introduction}
\label{introsec}

General Relativity (GR) is a consistent theory of gravity 
describing the gravitational law on Solar-System scales. 
On the other hand, the observational evidence of 
late-time cosmic acceleration \cite{obser1,obser2,obser3} 
suggests that one needs to introduce an unknown component 
dubbed dark energy in the context of GR. 
An alternative way of explaining the cosmic acceleration 
is to modify the gravitational law at large distances. 
Indeed, there have been many attempts for constructing 
models based on large-distance modifications of gravity, while 
recovering the behavior close to GR 
inside the Solar System \cite{review}.

If we turn our attention to the extreme short-distance or 
high-energy physics like the big bang and gravitational collapse,  
it is known that singularities inevitably arise in GR \cite{Hawking69}. 
In such strong gravitational regimes, we cannot exclude 
a possibility that GR is subject to modifications. 
In particular, after the detection of gravitational 
waves from 
black hole (BH) mergers \cite{GW1}, we are entering 
a golden era in which the 
physics of BHs and their surroundings can be observationally probed with increasing accuracy.
This will shed new light on the possible deviation from GR in the 
nonlinear regime of gravity.

In GR, the property of BHs is characterized by 
three ``hairs''--mass $M$, electric charge $Q$, and 
angular momentum $a$ \cite{nohair}. 
In theories beyond GR, the existence of additional 
degrees of freedom (DOFs) can give rise to new hairs 
to the field configuration and spacetime metric. 
The theories containing a scalar field $\phi$
coupled to gravity besides two tensor polarizations arising 
from the gravity sector are dubbed scalar-tensor 
theories \cite{Fujii}. 
In particular, Horndeski \cite{Horndeski} constructed most 
general scalar-tensor theories with second-order equations of motion. 
In shift-symmetric Horndeski theories invariant under the shift 
$\phi \to \phi+b$, where $b$ is a constant, 
there exists a no-hair theorem for static and spherically 
symmetric BHs based on the regularity of a Noether current 
on the horizon \cite{Hui}. It is however possible to realize a hairy BH solution for $\phi$ linearly coupled to 
a Gauss-Bonnet term \cite{Soti} 
by evading one of the conditions 
assumed in Ref.~\cite{Hui}.
If we allow for a time-dependence of $\phi$ or abandon
the shift symmetry, there are other hairy BHs 
arising in Horndeski theories \cite{timesca,Kanti} 
(see also Ref.~\cite{stensorBH}). 
The stability analyses of black holes in scalar-tensor theories 
were also performed in Refs.~\cite{odpapers1,modestability}. 

For a vector field coupled to gravity, it is known that 
generalized Proca theories \cite{Heisenberg,Tasinato,Allys,Jimenez} 
are the most general vector-tensor 
theories with second-order equations of motion. 
Apart from a specific intrinsic vector-mode coupling 
advocated by Horndeski in 1976 \cite{Horndeski76}, 
the $U(1)$ gauge 
invariance is explicitly broken by the presence of 
derivative interactions or nonminimal couplings to gravity.
The breaking of $U(1)$ gauge invariance leads to the propagation of a longitudinal scalar besides two 
transverse vector modes and two tensor polarizations. 
In vector-tensor theories, the existence of a temporal vector 
component gives rise to a bunch of hairy BH solutions \cite{GPBH1,GPBH2,GPBH3} without 
tunings of the models. 
The stability analysis against odd-parity perturbations 
on a static and spherically symmetric 
background \cite{KMTZ} 
shows that some BH solutions with nontrivial behavior of 
the longitudinal mode $A_1$ around the horizon are
excluded (including those found in Ref.~\cite{GPBH1}). 
A healthy extension of generalized Proca theories \cite{HKT} 
allows the possibility for evading the BH 
instability \cite{KMT}.

These two important classes of field theories, Horndeski and generalized Proca, can be unified
in the framework of scalar-vector-tensor (SVT) 
theories with second-order equations of 
motion \cite{Heisenberg18}. 
The SVT theories can be classified into two cases 
depending on whether they respect the $U(1)$ gauge 
symmetry or not. In the presence of $U(1)$ gauge 
symmetry the longitudinal component of a vector field 
vanishes, so that the propagating DOFs are 
five in total (one scalar, two transverse vectors, 
two tensor polarizations). The breaking of $U(1)$ gauge 
symmetry leads to the propagation of the longitudinal 
scalar besides the five DOFs. 
In the gauge-invariant case, two of the present authors 
found a new type of hairy BH solutions endowed with 
scalar and vector hairs in the presence of a cubic-order 
coupling \cite{HT18} (see also Refs.~\cite{Cha,Pedro}).
It remains to be seen whether such hairy BHs are stable 
against perturbations on the static and spherically 
symmetric background.

In this paper, we study the stability of static and spherically 
symmetric BHs against odd-parity perturbations in $U(1)$ 
gauge-invariant SVT theories. 
Since the analysis of even-parity perturbations is 
generally more involved, we leave the full
stability analysis against odd- and even-parity perturbations 
for a future work.
In Sec.~\ref{modelsec}, we first revisit gauge-invariant SVT theories and hairy BH solutions found in Ref.~\cite{HT18}.
In Sec.~\ref{stasec}, 
we will derive conditions for the absence of 
ghosts and Laplacian instabilities by expanding the 
most general action of gauge-invariant SVT theories 
up to second order in perturbations. 
Applying those conditions to concrete hairy BH 
solutions, we show in Sec.~\ref{stasec2} 
that the BH induced by the cubic-order coupling is 
stable against odd-parity perturbations under 
a certain bound of the coupling constant.  
As expected, the propagation speeds of 
perturbations arising from the gravity sector are 
equivalent to the speed of light in both radial and 
angular directions. 
If quartic-order couplings are present besides
the cubic-order coupling, we show that the BHs can be 
stable against odd-parity perturbations under certain conditions.
We conclude in Sec.~\ref{concludesec}.
Throughout the paper, we use the natural unit where 
the speed of light $c$ is equivalent to 1.

\section{Hairy black holes in gauge-invariant 
SVT theories}
\label{modelsec}

We consider the theories with $U(1)$ gauge-invariant 
SVT interactions with a scalar field $\phi$ 
and a vector field $A_{\mu}$. 
Besides these new interactions, we also take into account 
the Einstein-Hilbert term $M_{\rm pl}^2 R/2$ in the 
Lagrangian, where $M_{\rm pl}$ is the reduced Planck mass 
and $R$ is the Ricci scalar. 
Then, the theories we study are given by the 
action \cite{Heisenberg18}
\be
\mathcal{S}=\int d^4x \sqrt{-g}\left(\frac{M_{\rm pl}^2}{2}R
+\sum_{i=2}^4\mathcal{L}^i_{\rm SVT}\right)\,,
\label{action}
\ee
where $g$ is a determinant of the metric tensor 
$g_{\mu \nu}$, and 
\ba
\label{genLagrangianSVT}
\mathcal{L}^2_{\rm SVT}&=&f_2(\phi,X,F,\tilde{F},Y)\,, 
\label{L2}\\
\mathcal{L}^3_{\rm SVT}&=&
\left[ f_3(\phi,X)g_{\rho\sigma}+\tilde{f}_3(\phi,X)
\nabla_\rho \phi\nabla_\sigma \phi \right] \tilde{F}^{\mu\rho}\tilde{F}^{\nu\sigma} \nabla_\mu \nabla_\nu \phi\,, \\
\mathcal{L}^{4}_{\rm SVT}&=&
f_4(\phi,X)L^{\mu\nu\alpha\beta}F_{\mu\nu}F_{\alpha\beta}
+\left[ 
\frac12f_{4,X}(\phi,X)+\tilde{f}_4(\phi) \right] 
\tilde{F}^{\mu\nu}\tilde{F}^{\alpha\beta}
\nabla_\mu\nabla_\alpha \phi\nabla_\nu\nabla_\beta\phi\,.
\label{L4}
\ea
Here, $\nabla_{\mu}$ is the covariant derivative operator, 
and 
\ba
&&
X=-\frac{1}{2} \nabla_{\mu} \phi  \nabla^{\mu} \phi\,,
\qquad
F=-\frac{1}{4} F_{\mu \nu} F^{\mu \nu}\,,
\qquad
\tilde{F}=-\frac{1}{4} F_{\mu \nu} \tilde{F}^{\mu \nu}\,,
\qquad \notag\\
&&
F_{\mu \nu}=\nabla_{\mu}A_{\nu}-\nabla_{\nu}A_{\mu}\,,
\qquad
\tilde{F}^{\mu\nu}=\frac{1}{2}
\mathcal{E}^{\mu\nu\alpha\beta} F_{\alpha\beta}\,,
\qquad
Y=\nabla_{\mu}\phi \nabla_{\nu}\phi 
F^{\mu \alpha}{F^{\nu}}_{\alpha}\,,
\ea
with the anti-symmetric Levi-Civita tensor 
$\mathcal{E}^{\mu\nu\alpha\beta}$ satisfying the normalization $\mathcal{E}^{\mu\nu\alpha\beta}
\mathcal{E}_{\mu\nu\alpha\beta}=-4!$. 
The double dual Riemann tensor $L^{\mu \nu \alpha \beta}$ is defined by 
\be
L^{\mu\nu\alpha\beta}=\frac{1}{4}
\mathcal{E}^{\mu\nu\rho\sigma}
\mathcal{E}^{\alpha\beta\gamma\delta} R_{\rho\sigma\gamma\delta}\,,
\ee
where $R_{\rho\sigma\gamma\delta}$ is the Riemann tensor.
The function $f_2$ depends on $\phi, X, F, \tilde{F}, Y$, whereas 
$f_3, \tilde{f}_3, f_4$ are functions of $\phi,X$ 
with the notation $f_{4,X} \equiv \partial f_4/\partial X$.
The function $\tilde{f}_4$ depends on $\phi$ alone.
The dependence of $\tilde{F}$ and $Y$ in ${\cal L}_{\rm SVT}^2$ on 
a static and spherically symmetric background either vanishes or 
can be expressed in terms of $X$ and $F$ as $Y=4XF$.
Therefore, we shall not consider such dependence in the following. 

In Ref.~\cite{HT18}, it was shown that hairy BH solutions
exist on the static and  spherically symmetric 
background given by the line element 
\be
ds^2=-f(r) dt^{2} +h^{-1}(r) dr^{2}
+ r^{2} \left(d\theta^{2}+\sin^{2}\theta\,d\varphi^{2} 
\right)\,,
\label{metric_bg}
\ee
where $f$ and $h$ depend on the radial coordinate $r$. 
On the background (\ref{metric_bg}), the scalar field 
$\phi$ and the components of $A_{\mu}$ are 
functions of $r$, such that $\phi=\phi(r)$ and 
$A_{\mu}=(A_0(r), A_1(r), 0, 0)$ \cite{screening}. 
Since we are now considering the $U(1)$ gauge-invariant 
theory, the longitudinal mode $A_1(r)$ does 
not contribute to the vector-field dynamics. 
On the background (\ref{metric_bg}), the quantities $X$ 
and $F$ reduce, respectively, to 
$X=-h \phi'^2/2$ and $F=h A_0'^2/(2f)$, where 
a prime represents a derivative with respect to $r$.

The background equations of motion following from 
the variation of the action (\ref{action}) with respect to 
$f,h,\phi,A_0$ are given, respectively, by \cite{HT18}
\ba
M_{\rm pl}^2 rfh' &=& M_{\rm pl}^2 f(1-h)
+r^2 \left( f f_2-hA_0'^2 f_{2,F} \right)
-2r h^2 \phi' A_0'^2f_3+hA_0'^2 \{ 
4(h-1)f_4-h^2\phi'^2 (f_{4,X}+2\tilde{f}_4) \}\,,\label{be1}\\
M_{\rm pl}^2 rh f' &=& M_{\rm pl}^2 f(1-h)
+r^2 \left( f f_2+fh \phi'^2 f_{2,X}-h A_0'^2 f_{2,F} 
\right)-2rh^2 \phi'A_0'^2 \left(3f_3-h \phi'^2 f_{3,X} \right) 
\nonumber \\
& &+hA_0'^2 \left[ 4(3h-1)f_4-h (9h-4)\phi'^2f_{4,X} 
+h^3 \phi'^4 f_{4,XX}-10h^2\phi'^2 \tilde{f}_4
\right]\,,\label{be2}\\
J_{\phi}' &=& {\cal P}_{\phi}\,,\label{be3}\\
J_{A}' &=& 0\,,\label{be4}
\ea
where 
\ba
\hspace{-0.7cm}
J_{\phi} &=& -\sqrt{\frac{h}{f}} \left[  
r^2 f f_{2,X}\phi' -2h A_0'^2 (2h \tilde{f}_4+3hf_{4,X}-2f_{4,X})\phi' 
+2rh^2 A_0'^2f_{3,X}\phi'^2 
+h^3 A_0'^2 f_{4,XX}\phi'^3
-2rhA_0'^2 f_3 \right]\,,\label{Jphi}\\
\hspace{-0.7cm}
{\cal P}_{\phi} &=&
\frac{1}{\sqrt{fh}} \left[ r^2ff_{2,\phi}+hA_0'^2
\{ 4f_{4,\phi}+2h (r\phi'f_{3,\phi}-2f_{4,\phi})
+h^2 (f_{4,X\phi}+2\tilde{f}_{4,\phi})\phi'^2 
\} \right]\,,\\
\hspace{-0.7cm}
J_A &=& \sqrt{\frac{h}{f}} A_0' \left[ 
r^2 f_{2,F}+4rh\phi' f_3+8(1-h)f_4
+2h^2\phi'^2 (f_{4,X}+2\tilde{f}_4) \right]\,.
\label{JA}
\ea
The current $J_A$ is conserved due to the $U(1)$ gauge symmetry. 
The coupling $\tilde{f}_3$ does not appear in 
the background Eqs.~(\ref{be1})-(\ref{be4}) due to 
the underlying background symmetry.

In Ref.~\cite{HT18}, it was shown that hairy BH solutions 
exist for the theories given by the functions 
$f_2=X+F$ and $f_3=\beta_3$, 
where $\beta_3$ is a constant. 
Provided that the cubic coupling $f_3=\beta_3$ is present, 
there are also hairy BH solutions in the presence of 
quartic couplings $f_4=\beta_4 X^n$, where $\beta_4$ 
and $n~(\geq 0)$ are constants.
Consider the theories with the functions
\be
f_2=X+F\,,\qquad f_3=\beta_3\,, \qquad 
f_4=\beta_4\,,\qquad \tilde{f}_4=0\,.
\label{f24}
\ee
The event horizon is characterized by the radial distance 
$r_h$ satisfying $f(r_h)=0$ and $h(r_h)=0$. 
In the vicinity of the horizon, the iterative solutions
to Eqs.~(\ref{be1})-(\ref{be4}), expanded up to the 
order of $(r/r_h-1)^2$, are \cite{HT18}
\ba
\hspace{-0.9cm}
f &=&\left(1-\mu \right) \left( \frac{r}{r_h}-1 \right)
-\frac{1-2\mu+12 \tilde{\beta}_3^2 \mu^2 (1-\mu)
+4\tilde{\beta}_4 (24\tilde{\beta}_4 \mu^2-40\tilde{\beta}_4 
\mu+3\mu^2
+16\tilde{\beta}_4-9\mu+4)}{(1+8\tilde{\beta}_4)^2}
\left( \frac{r}{r_h}-1 \right)^2,\label{fho2}\\
\hspace{-0.9cm}
h &=& \left(1-\mu \right) \left( \frac{r}{r_h}-1 \right)
-\frac{1-2\mu-4 \tilde{\beta}_3^2 \mu^2 (1-\mu)
-4\tilde{\beta}_4 (8\tilde{\beta}_4 \mu^2
+8\tilde{\beta}_4 \mu+\mu^2
-16\tilde{\beta}_4+5\mu-4)}{(1+8\tilde{\beta}_4)^2}
\left( \frac{r}{r_h}-1 \right)^2\,,\label{hho2}\\
\hspace{-0.9cm}
A_0 &=& a_0+\sqrt{\frac{2\mu}{1+8\tilde{\beta}_4}}
M_{\rm pl}
\left( \frac{r}{r_h}-1 \right)
-\sqrt{\frac{2\mu}{(1+8\tilde{\beta}_4)^5}}M_{\rm pl} 
\left[ 1+4\tilde{\beta}_3^2 \mu (2-\mu) 
+4\tilde{\beta}_4-32\tilde{\beta}_4^2\right]
\left( \frac{r}{r_h}-1 \right)^2\,,
\label{A0ho2}\\
\hspace{-0.9cm}
\phi' &=& \frac{4 \tilde{\beta}_3\mu M_{\rm pl}}
{r_h(1+8\tilde{\beta}_4)} 
\left[ 1- \frac{5+32 \tilde{\beta}_3^2  \mu(1-\mu)
+16\tilde{\beta}_4 (2+\mu-4\tilde{\beta}_4
+8\tilde{\beta}_4 \mu)}{(1+8\tilde{\beta}_4)^2}
\left( \frac{r}{r_h}-1 \right) \right],
\label{phiso1}
\ea
where $\tilde{\beta}_3=\beta_3 M_{\rm pl}/r_h^2$, 
$\tilde{\beta}_4=\beta_4/r_h^2$, and $\mu$ is a constant 
in the range $0<\mu<1$. In the above expressions, 
we have chosen the branch $A_0'>0$ at $r=r_h$.
For $\tilde{\beta}_3 \neq 0$, 
there is a nonvanishing scalar hair ($\phi' \neq 0$).
The couplings $\tilde{\beta}_3$ and $\tilde{\beta}_4$ lead
to modifications to the metric components 
$f_{\rm RN}=h_{\rm RN}=(1-r_h/r)(1-\mu r_h/r)$ 
and the temporal vector component 
$A_0^{\rm RN}=P+Q/r$ of the 
Reissner-Nordstr\"{o}m (RN) solution ($P$ and $Q$ 
are constants). 
At spatial infinity ($r \gg r_h$), the iterative solutions, 
up to the order of $1/r^8$, are given by 
\ba
f&=& 1-\frac{2M}{r}+\frac{Q^2}{2M_{\rm pl}^2 r^2}
-\frac{2\beta_4 Q^2}{M_{\rm pl}^2r^4}
+\frac{2\beta_4 MQ^2}{M_{\rm pl}^2r^5}
-\frac{3\beta_4 Q^4}{5M_{\rm pl}^4r^6}
+\frac{256\beta_4^2 MQ^2}{7M_{\rm pl}^2r^7}
\nonumber \\
& &
+\frac{3Q^2 (M_{\rm pl}^2Q^2 \beta_3^2 
-28\beta_4^2 Q^2-256\beta_4^2 M^2M_{\rm pl}^2)}{14M_{\rm pl}^4 r^8}\,,
\label{fho2d}\\
h&=& 1-\frac{2M}{r}+\frac{Q^2}{2M_{\rm pl}^2 r^2}
-\frac{2\beta_4 M Q^2}{M_{\rm pl}^2 r^5}
+\frac{2\beta_4 Q^4}{5M_{\rm pl}^4 r^6}
-\frac{2Q^2 (\beta_3^2 Q^2-64\beta_4^2M^2)}
{7M_{\rm pl}^2 r^8}\,,\\
A_0 &=& P+\frac{Q}{r}-\frac{4\beta_4M Q}{r^4}
+\frac{3\beta_4 Q^3}{5M_{\rm pl}^2r^5}
-\frac{8Q (\beta_3^2 Q^2-32\beta_4^2M^2)}{7r^7}
+\frac{2MQ^3 (7\beta_3^2 M_{\rm pl}^2
-48 \beta_4^2)}{7M_{\rm pl}^2r^8}\,,
\label{A0so2}\\
\phi' &=& \frac{2\beta_3 Q^2}{r^5}
-\frac{64 \beta_3 \beta_4 MQ^2}{r^8}\,,
\label{phiso2}
\ea
where $M$ is a constant. 
Again, the coupling $\beta_3$ induces a nonvanishing 
scalar hair. The RN solution with 
$f=h=1-2M/r+Q^2/(2M_{\rm pl}^2r^2)$ and 
$A_0=P+Q/r$ is subject to modifications by 
the couplings $\beta_3$ and $\beta_4$. 
Due to the current conservation (\ref{be4}), the $U(1)$ charge 
$Q$ at spatial infinity is related to the quantities $\mu$ and 
$r_h$ in the vicinity of the horizon, as 
$\sqrt{2\mu (r_h^2+8\beta_4)}M_{\rm pl}=-Q$. 
Solving Eqs.~(\ref{be1})-(\ref{be4}) with the functions (\ref{f24}) numerically, 
the iterative solutions (\ref{fho2})-(\ref{phiso1}) 
around the horizon smoothly connect to the solutions (\ref{fho2d})-(\ref{phiso2}) at spatial 
infinity \cite{HT18}.
Thus, there are regular BHs endowed with 
scalar and vector hairs.

Provided that the cubic-coupling $f_3=\beta_3$ 
is present, there are also hairy BH solutions for 
quartic-order power  
couplings $f_4=\beta_4 X^n$ with $n \geq 1$.
For $n=1$, the iterative solutions to 
$f, h, A_0, \phi'$ in the vicinity of 
the horizon are given by Eqs.~(4.15)-(4.18) 
of Ref.~\cite{HT18}.
At spatial infinity, the solutions to $f, h, A_0$, 
for $n=1$, are expressed in 
the forms (3.19)-(3.21) of Ref.~\cite{HT18} up 
to the order of $1/r^8$, with the leading-order 
scalar derivative $\phi'=2\beta_3 Q^2/r^5$.

\section{General BH stability against odd-parity perturbations }
\label{stasec}

Let us consider small perturbations $h_{\mu \nu}$ on top of 
the static and spherically symmetric background (\ref{metric_bg}). For the study of odd-parity 
perturbations we choose the Regge-Wheeler gauge 
$h_{ab}=0$ \cite{Regge:1957td,Zerilli:1970se}, 
where $a,b$ represent either $\theta$ 
or $\varphi$. Then, the metric perturbations corresponding 
to odd-mode perturbations are expressed 
in the form \cite{odpapers1,odpapers2}
\be
h_{tt}=h_{tr}=h_{rr}=0\,,\qquad
h_{ta}=\sum_{l,m}Q_{lm}(t,r)E_{ab}
\partial^bY_{lm}(\theta,\varphi)\,,
\qquad
h_{ra}=\sum_{l,m}W_{lm}(t,r)E_{ab}\partial^bY_{lm}(\theta,\varphi)\,,
\ee
where $Q_{lm}$ and $W_{lm}$ are functions 
of $t$ and $r$, and $Y_{lm}(\theta, \varphi)$ is 
the spherical harmonics.
The tensor $E_{ab}$ is given by 
$E_{ab}=\sqrt{\gamma}\, \varepsilon_{ab}$, where 
$\gamma$ is the determinant of the metric 
$\gamma_{ab}$ on the two-dimensional sphere and 
$\varepsilon_{ab}$ is the anti-symmetric symbol 
with $\varepsilon_{\theta\varphi}=1$. 
The scalar field $\phi$ does not have odd-parity 
perturbations. The perturbations of $A_{\mu}$ 
for the odd-parity sector are given by \cite{KMTZ}
\be
\da_{t}=\da_{r}=0\,,\qquad 
\da_{a}=\sum_{l,m}\da_{lm}(t,r)E_{ab}\partial^bY_{lm}(\theta,\varphi)\,,
\ee
where $\da_{lm}$ depends on $t$ and $r$.

\subsection{Second-order action}

We expand the action (\ref{action}) up to second order 
in odd-parity perturbations. 
In doing so, we can set $m=0$ without loss of generality.
The integrations with respect to $\theta$ and $\varphi$ are performed by using the properties of spherical harmonics given 
in Appendix B of Ref.~\cite{KMTZ}. 
We also integrate the action by parts with respect to 
$t, r$ and finally employ the background 
Eqs.~(\ref{be1}), (\ref{be2}), and (\ref{be4}) to eliminate 
the terms $f_2, f_{2,X}, f_{3,X}$.
Then, the second-order action of odd-parity 
perturbations yields 
\ba
{\cal S}_{\rm odd}^{(2)}&=&\sum_{l,m} L \int dt dr\, 
{\cal L}_{\rm odd}^{(2)}\,, 
\label{oddact}
\ea
where $L=l(l+1)$, and  
\ba
{\cal L}_{\rm odd}^{(2)}&=&
r^2 \sqrt{\frac{f}{h}}
\biggl[
\alpha_1\left(\dot{W}_{lm}-Q'_{lm}+\frac{2}{r}Q_{lm}\right)^2+2\left( \alpha_2 \da'_{lm}+\alpha_3 \da_{lm} \right)
\left(\dot{W}_{lm}-Q'_{lm}+\frac{2}{r}Q_{lm}\right)
+\alpha_4 \dot{\da}_{lm}^2
\notag\\
&&
+\alpha_5 {\da}_{lm}'^2+(L-2) \left( \alpha_6 W_{lm}^2
+\alpha_7 Q_{lm}^2 
+\alpha_8Q_{lm}\da_{lm} \right)
+L\alpha_9 \da_{lm}^2 \biggr]\,.
\label{oddLag}
\ea
Here, a dot represents the derivative with respect to $t$, 
and the coefficients $\alpha_i$ are given by
\ba
& &
\alpha_1=\frac{M_{\rm pl}^2h}{4fr^2}\,,\qquad 
\alpha_2=\frac{h^2A_0'}{2fr^3} \left[ 
r\phi' f_3-4f_4+h\phi'^2 (f_{4,X}+2\tilde{f}_4) \right]\,,
\nonumber \\
& &
\alpha_3=-\frac{hA_0'}{2fr^4} 
\left[ f_{2,F}\,r^2+4h (r \phi' f_3-2f_4)+2h^2 
\phi'^2 (f_{4,X}+2\tilde{f}_4) \right]\,,
\nonumber \\
& &
\alpha_4=\frac{1}{2fr^3} \left[ f_{2,F}\,r+ 
( 2rh \phi''+rh' \phi'+2h \phi')f_3 +2h^2 \phi'^3 \tilde{f}_3
-4h' f_4+h \phi' (2h \phi''+h'\phi') (f_{4,X}+2\tilde{f}_4)
\right]\,,
\nonumber \\
& &
\alpha_5=-\frac{h}{2fr^3} \left[ f f_{2,F}\,r
+h \phi' (2f+f' r)f_3-4f'h f_4+f'h^2 \phi'^2
(f_{4,X}+2\tilde{f}_4) \right]\,,
\nonumber \\
& &
\alpha_6=-\frac{h}{4fr^4} 
\left[ M_{\rm pl}^2 f-4h A_0'^2 f_4
+h^2 \phi'^2 A_0'^2 (f_{4,X}+2\tilde{f}_4) \right]\,,
\qquad
\alpha_7=\frac{M_{\rm pl}^2 f-4h A_0'^2f_4}
{4f^2r^4}\,,
\nonumber \\
& &
\alpha_8=\frac{2}{f^2 r^4} \left[ (2fh A_0''+fh' A_0'
-h f' A_0')f_4 -f h A_0' \phi' (2h \phi''+h'\phi')f_{4,X}
+2fh \phi' A_0' f_{4,\phi} \right]\,,
\nonumber \\
& &
\alpha_9=-\frac{1}{4f^2r^4} \biggl[ 2f^2 f_{2,F} 
+2f (2 fh \phi''+f' h \phi'+f h' \phi')f_3+2f f' h^2 \phi'^3 
\tilde{f}_3+4 (f'^2 h -f f' h'-2ff'' h)f_4 \nonumber \\
& &
\qquad \qquad \qquad~+f f' h\phi' (2h \phi''+h' \phi')(f_{4,X}
+2\tilde{f}_4) \biggr]\,.
\label{alphai}
\ea

\subsection{Dipole perturbations ($l=1$)}

We first consider the dipole mode $l=1$, i.e., $L=2$. 
Since the perturbations $h_{ab}$ identically vanish  
for $l=1$, we cannot choose the Regge-Wheeler gauge.
Under the gauge transformation 
$x_\mu \to x_\mu+\xi_\mu$, 
where $\xi_t=\xi_r=0$ and $\xi_a=
\sum_{l,m} \Lambda_{lm}(t,r) E_{ab}\partial^b 
Y_{lm}(\theta, \varphi)$ for odd-parity modes,  
the perturbations $Q_{lm}$ and $W_{lm}$ 
transform, respectively, to 
\be
Q_{lm} \to Q_{lm}+\dot{\Lambda}_{lm}\,,\qquad 
W_{lm} \to W_{lm}+\Lambda_{lm}'-\frac{2}{r}\Lambda_{lm}\,.
\ee
For the dipole mode, we choose the gauge 
\be
W_{1m}=0\,,
\label{W1m}
\ee
under which the quantity $\Lambda_{1m}$ in $\xi_a$ 
is given by 
\be
\Lambda_{1m} (t,r)=-r^2\int d\tilde{r}
\frac{W_{1m}(t,\tilde{r})}{\tilde{r}^2}
+r^2{\cal C}(t)\,, 
\label{Lam1m}
\ee
where ${\cal C}(t)$ is an arbitrary function of $t$.
We note that the terms proportional to $L-2$ in the 
Lagrangian (\ref{oddLag}) vanish for dipole perturbations. 
Varying the action (\ref{oddact}) with respect to $W_{1m}$ and $Q_{1m}$ and finally using the gauge condition (\ref{W1m}), we obtain 
\be
\dot{\cal E}=0\,,\qquad 
\left( r^2 {\cal E} \right)'=0\,,
\label{Eeq}
\ee
where 
\be
{\cal E}=r^2 \sqrt{\frac{f}{h}} 
\left[ \alpha_1 \left( Q_{1m}'-\frac{2}{r} Q_{1m} 
\right)-\left( \alpha_2 \delta A_{1m}'+\alpha_3 
\delta A_{1m} \right) \right]\,.
\ee
The solution to Eq.~(\ref{Eeq}) is given by 
${\cal E}={\cal C}_1/r^2$, where ${\cal C}_1$ is a constant. 
Then, it follows that 
\be
\alpha_1 \left( Q_{1m}'-\frac{2}{r} Q_{1m} 
\right)=\alpha_2 \delta A_{1m}'+\alpha_3 
\delta A_{1m}+\frac{{\cal C}_1}{r^4}
\sqrt{\frac{h}{f}}\,,
\label{Q1m}
\ee
which can be written in the integrated form 
\be
Q_{1m}=r^2 \int d \tilde{r} \frac{1}{\alpha_1 \tilde{r}^2} 
\left( \alpha_2 \delta A_{1m}'+\alpha_3 
\delta A_{1m}+\frac{{\cal C}_1}{\tilde{r}^4}
\sqrt{\frac{h}{f}} \right)+r^2 {\cal C}_2(t)\,,
\ee
where ${\cal C}_2(t)$ is an arbitrary function of $t$. 
The residual gauge degree of freedom ${\cal C}(t)$ in 
Eq.~(\ref{Lam1m}) can be fixed by choosing 
${\cal C}(t)=\int d\tilde{t}\,{\cal C}_2 (\tilde{t})$.

On using Eq.~(\ref{Q1m}) to eliminate the combination 
$ Q_{1m}'-2Q_{1m}/r$ from Eq.~(\ref{oddLag}), 
the second-order Lagrangian (\ref{oddLag}) yields
\be
{\cal L}_{\rm odd}^{(2)}
=r^2 \sqrt{\frac{f}{h}} \left[ \alpha_4 \dot{\da}_{1m}^2
+\left( \alpha_5 -\frac{\alpha_2^2}{\alpha_1} \right)
\da_{1m}'^2-\frac{2\alpha_2 \alpha_3}{\alpha_1}
\da_{1m}\da_{1m}'+\left(2\alpha_9-\frac{\alpha_3^2}
{\alpha_1} \right) \da_{1m}^2
+\frac{h\,{\cal C}_1^2}{\alpha_1 fr^8} \right]\,.
\label{Lodd2}
\ee
This shows that the vector-field perturbation $\da_{1m}$ 
is the only propagating DOF for dipole perturbations.
The ghost is absent as long as the first term in the square bracket of Eq.~(\ref{Lodd2}) is positive, i.e.,  
\be
\alpha_4>0\,.
\ee

In Fourier space, we consider the solution to the 
vector-field perturbation in the form 
$\da_{1m} \propto e^{i (\omega t-kr)}$, 
where $\omega$ is a frequency and $k$ is a comoving wavenumber. In the small-scale limit, the dominant contributions to ${\cal L}_{\rm odd}^{(2)}$ are the first 
two terms in the square brackets of Eq.~(\ref{Lodd2}).  
Then, the dispersion relation corresponds to 
$\alpha_4 \omega^2+(\alpha_5-\alpha_2^2/\alpha_1)k^2=0$. 
The speed of the perturbation $\da_{1m}$ along the radial direction 
in proper time is given by $\hat{c}_r=dr_*/d\tau$, where 
$dr_*=dr/\sqrt{h}$ and $d\tau=\sqrt{f}dt$.
This is related to the propagation speed $c_r=dr/dt$ 
in the coordinates $t$ and $r$, as $\hat{c}_r=\sqrt{fh}\,c_r$, where $\omega=\hat{c}_rk$. 
{}From the dispersion relation in the small-scale limit, 
we obtain 
\be
c_r^2=\frac{\alpha_2^2-\alpha_1 \alpha_5}{fh\,\alpha_1 \alpha_4}\,.
\label{crdipole}
\ee
We require the condition $c_r^2\geq0$ for the absence of Laplacian instabilities of vector-field perturbations in the odd-parity sector.

\subsection{Perturbations with $l \geq 2$}

Let us proceed to the discussion of stability conditions for 
odd-parity perturbations with $l \geq 2$. 
In the Lagrangian (\ref{oddLag}), there are two 
dynamical fields $W_{lm}$ and $\da_{lm}$, while
the field $Q_{lm}$ is non-dynamical.
To study the propagation of dynamical DOFs, 
it is convenient to rewrite the Lagrangian 
(\ref{oddLag}) in terms of a Lagrangian multiplier 
$\chi(t,r)$, as
\ba
\hspace{-0.8cm}
{\cal L}_{\rm odd}^{(2)}&=&r^2 \sqrt{\frac{f}{h}}
\left[ \alpha_1\left\{2\chi\left(\dot{W}_{lm}-Q'_{lm}
+\frac{2}{r}Q_{lm}+\frac{\alpha_2\da'_{lm}
+\alpha_3 \da_{lm}}{\alpha_1}\right)-\chi^2\right\}
-\frac{(\alpha_2\da'_{lm}+\alpha_3\da_{lm})^2}
{\alpha_1}
\right.
\notag\\
\hspace{-0.8cm}
&&\left.
+\alpha_4 \dot{\da}_{lm}^2
+\alpha_5 {\da}_{lm}'^2+(L-2) \left( \alpha_6 W_{lm}^2
+\alpha_7 Q_{lm}^2 
+\alpha_8Q_{lm}\da_{lm} \right)
+L\alpha_9 \da_{lm}^2\right]\,,
\label{LM}
\ea
whose variation with respect to $\chi$ 
leads to
\be
\chi=
\dot{W}_{lm}-Q'_{lm}
+\frac{2}{r}Q_{lm}+\frac{\alpha_2\da'_{lm}
+\alpha_3 \da_{lm}}{\alpha_1}\,.
\label{chi}
\ee
Substituting Eq.~(\ref{chi}) into Eq.~(\ref{LM}), we recover 
the original second-order Lagrangian (\ref{oddLag}).
Varying the Lagrangian (\ref{LM}) with respect to 
$W_{lm}$ and $Q_{lm}$, respectively, we obtain 
\ba
&&
\alpha_1\dot{\chi}-(L-2)\alpha_6 W_{lm}=0\,,
\label{eqW}\\
&&
\alpha_1 \chi'
+\frac{(8fh+rf'h-rfh')\alpha_1+2rfh\alpha_1'}{2rfh}\chi
+(L-2)\left( \alpha_7 Q_{lm}+\frac{\alpha_8}{2}\da_{lm} \right)=0\,. 
\label{eqQ}
\ea
We solve Eqs.~(\ref{eqW}) and (\ref{eqQ}) for 
$W_{lm}$ and $Q_{lm}$ respectively and substitute 
them into Eq.~(\ref{LM}). After integrations by parts, 
the second-order Lagrangian 
is expressed in the form 
\be
(L-2){\cal L}_{\rm odd}^{(2)}=r^2 \sqrt{\frac{f}{h}}\left( 
\dot{\vec{\mathcal{X}}}^{t}{\bm K}\dot{\vec{\mathcal{X}}}
+\vec{\mathcal{X}}'^{t}{\bm G}\vec{\mathcal{X}}'
+\vec{\mathcal{X}}'^{t}{\bm S}\vec{\mathcal{X}}
+\vec{\mathcal{X}}^{t}{\bm M}\vec{\mathcal{X}}
\right)\,,
\label{LM2}
\ee
where ${\bm {K,G,S,M}}$ are $2\times2$ matrices, with the vector 
\be
\vec{\mathcal{X}}^{t}=\left(\chi,\da_{lm} \right)\,.
\ee
This shows that there are two dynamical fields $\chi$ and 
$\da_{lm}$. The field $\chi$ arises from 
perturbations in the gravity sector (i.e., tensor modes), 
whereas the field $\da_{lm}$ corresponds to the vector 
degree of freedom. 
As we already mentioned, the perturbation of scalar field 
$\phi$ does not arise as a dynamical degree of freedom 
for odd-parity perturbations.
The nonvanishing components of the matrices 
${\bm {K,G,S,M}}$ are given by 
\ba
& &
K_{11}=-\frac{\alpha_1^2}{\alpha_6}\,,\qquad 
K_{22}=(L-2)\alpha_4\,,\qquad 
G_{11}=-\frac{\alpha_1^2}{\alpha_7}\,,\qquad 
G_{22}=\frac{(L-2)(\alpha_1 \alpha_5-\alpha_2^2)}{\alpha_1}\,, \nonumber \\
& &
S_{12}=-S_{21}=-(L-2)\left(\alpha_2+\frac{\alpha_1\alpha_8}{2\alpha_7}\right)\,,
\notag\\
&&
M_{11}=-(L-2)\alpha_1-\frac{h}{fr^8\alpha_7}
\left[\left(r^4\sqrt{\frac{f}{h}}\alpha_1\right)'\, \right]^2
+\frac{1}{2r^2}\sqrt{\frac{h}{f}}\left[\frac{r\alpha_1\{(f'hr-h'fr+8fh)\alpha_1
+2fhr \alpha_1'\}}{\sqrt{f}h^{3/2} \alpha_7}\right]'\,,
\notag\\
&&
M_{22}=-(L-2)\left[\frac{(L-2)\alpha_8^2}{4\alpha_7}-L\alpha_9
+\frac{\alpha_3^2}{\alpha_1}-\frac{1}{r^2}\sqrt{\frac{h}{f}}\left(r^2
\sqrt{\frac{f}{h}}\frac{\alpha_2\alpha_3}{\alpha_1}\right)'\,\right]\,,
\notag\\
&&
M_{12}=M_{21}=(L-2)\left[\alpha_3-\frac{1}{2r^4}\sqrt{\frac{h}{f}}
\frac{\alpha_8}{\alpha_7}\left(r^4\sqrt{\frac{f}{h}}\alpha_1\right)'
-\frac{1}{2r^2}\sqrt{\frac{h}{f}}\left\{r^2\sqrt{\frac{f}{h}}
\left(\alpha_2-\frac{\alpha_1\alpha_8}{2\alpha_7}\right)\right\}'\,
\right]\,.
\ea

Since there are no off-diagonal components for the 
matrix ${\bm K}$, the no-ghost conditions  
correspond to $K_{11}>0$ and $K_{22}>0$, i.e., 
\be
\alpha_6<0\,,\qquad \alpha_4>0\,. 
\label{noghost}
\ee

Let us consider the propagation of perturbations along the 
radial direction by assuming the solution of the form 
$\vec{\mathcal{X}}^{t} \propto e^{i (\omega t-kr)}$. 
In the small-scale limit ($k \to \infty$), 
the dispersion relation is expressed as 
${\rm det} \left( \omega^2 {\bm K}+k^2{\bm G} 
\right)=0$. The propagation speed $c_r$ 
in proper time can be derived by substituting $\omega=\sqrt{fh}\,c_r k$ 
into the dispersion relation. Then, we obtain the following 
two expressions of $c_r^2$ :
\ba
c_{r1}^2
&=& -\frac{G_{11}}{fh K_{11}}
=-\frac{\alpha_6}{fh\,\alpha_7}\,,
\label{cr1}\\
c_{r2}^2
&=& -\frac{G_{22}}{fh K_{22}}
=\frac{\alpha_2^2-\alpha_1 \alpha_5}
{fh\,\alpha_1 \alpha_4}\,.
\label{cr2}
\ea
We recall that, for dipole perturbations ($l=1$), only the 
vector-field perturbation $\da_{1m}$ is dynamical 
with the propagation speed squared $c_r^2$ given by 
Eq.~(\ref{crdipole}). This is equivalent to $c_{r2}^2$ 
derived above, which corresponds to the propagation 
speed squared of vector-field perturbations. 
The other value $c_{r1}^2$ is related to the propagation 
speed squared arising from the gravity sector. 
To avoid small-scale Laplacian instabilities 
along the radial direction, 
we require the two conditions 
$c_{r1}^2 \geq 0$ and $c_{r2}^2 \geq 0$.

In the limit that $L=l(l+1) \gg 1$, the matrix ${\bm M}$ contributes to the propagation speed $c_{\Omega}$ along the angular direction. In this limit, the matrix components 
$M_{11}$ and $M_{22}$ are given, respectively, by 
\be
M_{11} \simeq -L\alpha_1\,,\qquad 
M_{22} \simeq \frac{L^2 (4\alpha_7 \alpha_9
-\alpha_8^2)}{4\alpha_7}\,.
\ee
The off-diagonal components $M_{12}$ and $M_{21}$ also 
contain the term proportional to $L$, but their contributions 
to $c_{\Omega}$ can be neglected for $L \gg 1$. This is also the case for the 
matrix components of ${\bm S}$. 
Assuming the solution of the form 
$\vec{\mathcal{X}}^{t} \propto e^{i (\omega t-l \theta)}$, 
the dispersion relation is given by 
${\rm det}(\omega^2{\bm K}+{\bm M})=0$.
In proper time, the propagation speed along the angular
direction is $c_{\Omega}=\hat{c}_{\Omega}/\sqrt{f}$, 
where $\hat{c}_{\Omega}=rd\theta/dt$.
Substituting $\omega^2=\hat{c}_{\Omega}^2 l^2/r^2=
c_{\Omega}^2f l^2/r^2$ into the dispersion relation 
and solving it for $c_{\Omega}^2$ with the limit 
$L \gg 1$, we obtain the following two expressions of 
$c_{\Omega}^2$ :
\ba
c_{\Omega 1}^2&=&
-\frac{r^2 M_{11}}{l^2 f K_{11}}
=-\frac{r^2 \alpha_6}{f \alpha_1}\,,
\label{co1}\\
c_{\Omega 2}^2&=&
-\frac{r^2 M_{22}}{l^2 f K_{22}}
=\frac{r^2 (\alpha_8^2-4\alpha_7 \alpha_9)}{4f\alpha_4 \alpha_7}\,,
\label{co2}
\ea
which correspond to the propagation speed squares 
arising from the gravity sector and the vector-field 
perturbation,  respectively.
We require the two conditions $c_{\Omega 1}^2 \geq 0$ 
and $c_{\Omega 2}^2 \geq 0$ to avoid Laplacian 
instabilities along the angular direction.

Substituting the explicit forms of coefficients 
$\alpha_1, \alpha_6, \alpha_7$ into Eqs.~(\ref{cr1}) 
and (\ref{co1}), it follows that 
\ba
c_{r1}^2 &=& 1+\frac{h^2A_0'^2 \phi'^2 
(f_{4,X}+2\tilde{f}_4)}
{M_{\rm pl}^2f -4h A_0'^2 f_4}\,,\label{cr1s} \\
c_{\Omega1}^2 &=& 1+\frac{hA_0'^2
[h\phi'^2  (f_{4,X}+2\tilde{f}_4)-4f_4]}
{M_{\rm pl}^2 f}\,.
\label{co1s}
\ea
For the theories containing couplings up to cubic order, 
$c_{r1}^2=1$ and $c_{\Omega1}^2=1$. 
The presence of quartic-order couplings $f_4$ and 
$\tilde{f}_4$ generally leads to the values of 
$c_{r1}^2$ and $c_{\Omega1}^2$ different from 1.

\section{Odd-parity stability of hairy black holes}
\label{stasec2}

We apply general stability conditions derived in 
Sec.~\ref{stasec} to concrete models with hairy BH solutions. 
Let us focus on models given by the functions
\be
f_2=X+F\,,\qquad f_3=\beta_3\,, \qquad 
f_4=\beta_4X^n\,,\qquad \tilde{f}_4=0\,,
\label{conmodel}
\ee
where $\beta_3, \beta_4, n~(\geq 0)$ are constants. 
In the following, we will study the three different cases: 
(A) $\beta_4=0$, (B) $\beta_4 \neq 0$, $n=0$, and 
(C) $\beta_4 \neq 0$, $n=1$, in turn.

\subsection{$\beta_4=0$}
\label{caseA}

In this case, we have $f_4=0$ and $\tilde{f}_4=0$ in 
Eqs.~(\ref{alphai}), (\ref{cr1s}), and (\ref{co1s}). 
Then, it follows that 
\be
\alpha_6=-\frac{hM_{\rm pl}^2}{4r^4}\,,
\qquad c_{r1}^2=1\,,\qquad 
c_{\Omega 1}^2=1\,.
\label{alpha6}
\ee
The no-ghost condition $\alpha_6<0$ is satisfied 
outside the horizon ($h>0$). The propagation speed 
squares $c_{r1}^2$ and $c_{\Omega 1}^2$ are the same as 
those in GR. This means that the quadratic and cubic 
couplings do not affect the stability conditions 
of perturbations in the gravity sector as one would expect.

Let us investigate the odd-parity stability associated 
with the vector-field perturbation $\delta A_{lm}$. 
In the vicinity of the event horizon of hairy BHs, 
we resort to the iterative solutions (\ref{fho2})-(\ref{phiso1}) 
with $\beta_4=0$. 
Then, the quantities $\alpha_4, c_{r2}^2, c_{\Omega2}^2$ 
reduce, respectively, to 
\ba
\alpha_4 &=& \frac{1+4\mu (1-\mu)\tilde{\beta}_3^2}
{2r_h^2 (1-\mu)}\left(\frac{r}{r_h}-1\right)^{-1}
+{\cal O}((r/r_h-1)^{0})\,,\\
c_{r2}^2 &=& 1+\frac{8\mu (1-\mu) \tilde{\beta}_3^2
[5+32\mu (1-\mu) \tilde{\beta}_3^2]}
{1+4\mu (1-\mu)\tilde{\beta}_3^2}
\left(\frac{r}{r_h}-1\right)+{\cal O}((r/r_h-1)^{2})\,,
\label{cr2es}\\
c_{\Omega 2}^2 &=& 
\frac{1+8\mu (1-\mu)\tilde{\beta}_3^2}
{1+4\mu (1-\mu)\tilde{\beta}_3^2}
+{\cal O}(r/r_h-1)\,.
\label{cO2es}
\ea
Since the constant $\mu$ is in the range $0<\mu<1$, 
all the stability conditions $\alpha_4>0$, $c_{r2}^2 \geq 0$, 
and $c_{\Omega 2}^2 \geq 0$ hold around $r=r_h$ 
for arbitrary couplings $\tb_3$. 
In the limit that $r \to r_h$, $c_{r2}^2$ approaches 1, 
whereas $c_{\Omega2}^2$ approaches a constant 
different from 1. 

At spatial infinity, the background solutions are given by 
Eqs.~(\ref{fho2d})-(\ref{phiso2}) with $\beta_4=0$. 
Then, it follows that 
\ba
\alpha_4 &=& \frac{1}{2r^2}+\frac{M}{r^3}+
+{\cal O} \left( \frac{1}{r^4} \right)\,,
\label{al4} \\
c_{r2}^2 &=& 1+\frac{20\beta_3^2 Q^2}{r^6}
+{\cal O} \left( \frac{1}{r^7} \right)\,,
\label{cr2d} \\
c_{\Omega 2}^2 &=& 
1-\frac{4\beta_3^2 Q^2}{r^6}
+{\cal O} \left( \frac{1}{r^7} \right)\,,
\label{cW2}
\ea
and hence the stability conditions $\alpha_4>0$, $c_{r2}^2 \geq 0$, and $c_{\Omega 2}^2 \geq 0$ are trivially satisfied.

\begin{figure}[h]
\begin{center}
\includegraphics[height=3.3in,width=3.3in]{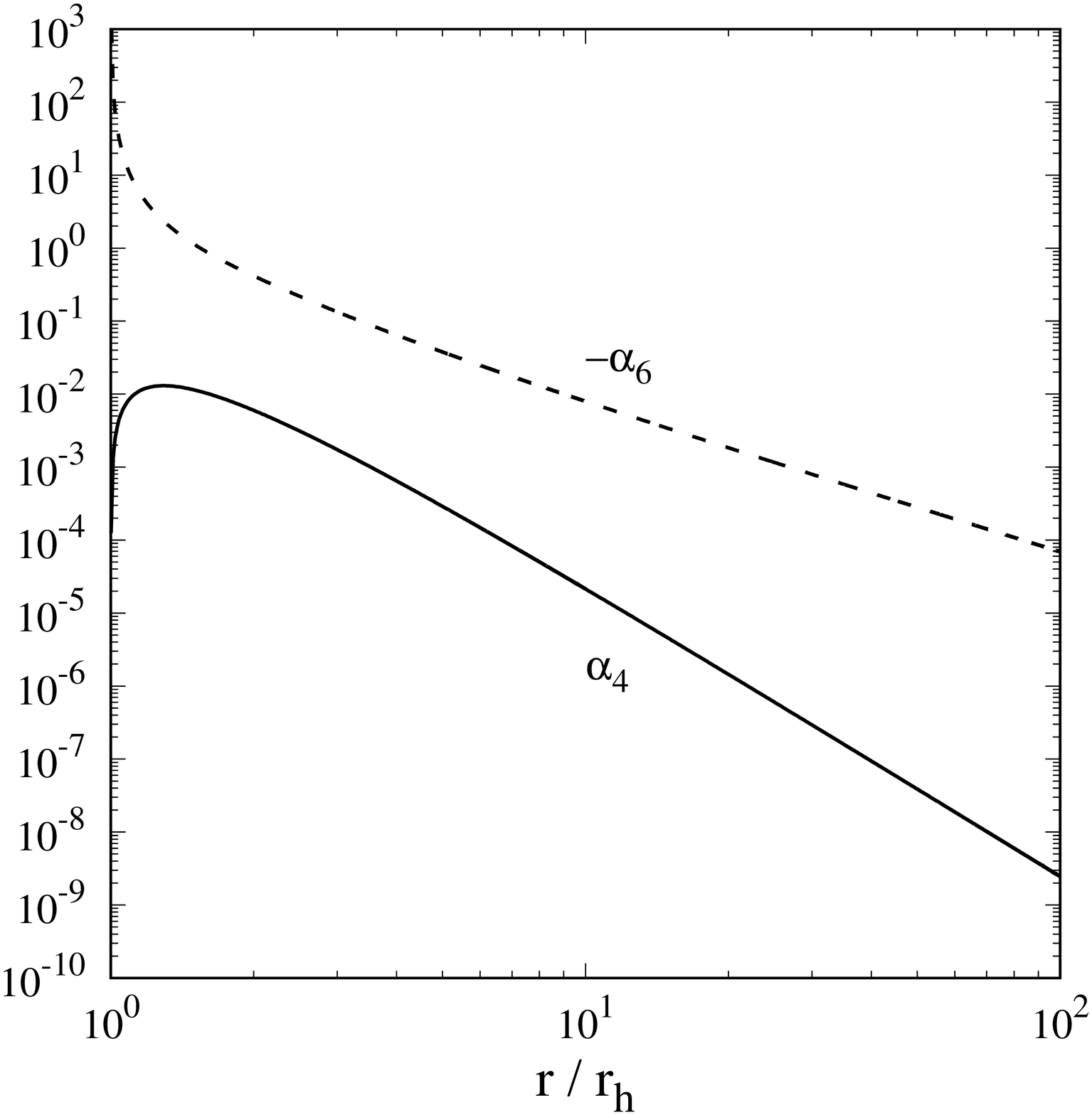}
\includegraphics[height=3.3in,width=3.3in]{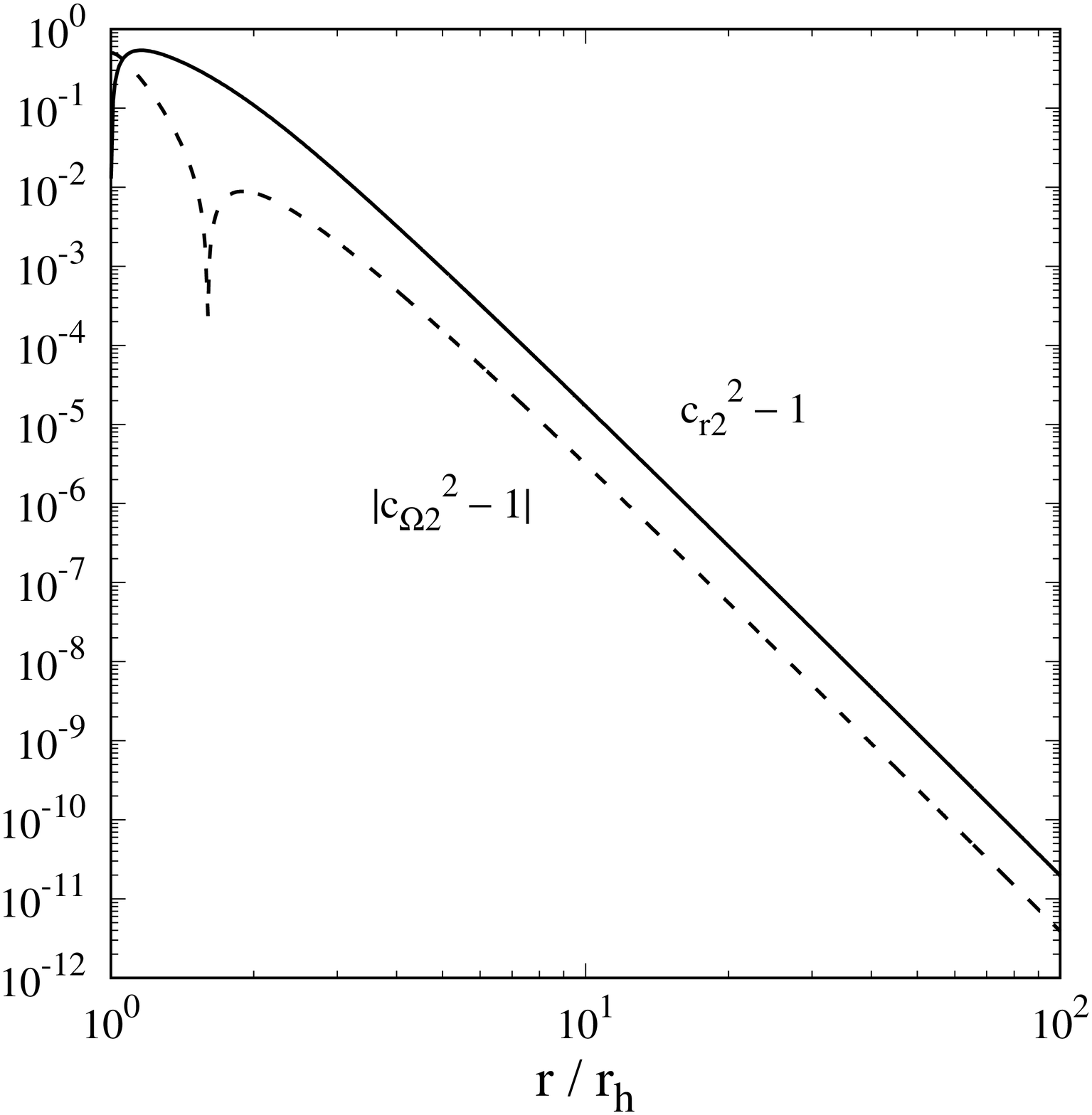}
\end{center}
\caption{\label{fig1}
Numerical solutions to $-\alpha_6, \alpha_4$ (left)
and $c_{r2}^2-1$, $|c_{\Omega_2}^2-1|$ (right) 
versus $r/r_h$
for the model (\ref{conmodel}) with 
$\tilde{\beta}_3=1$, $\beta_4=0$, and  $\mu=0.5$.
The background boundary conditions are chosen to 
be consistent with Eqs.~(\ref{fho2})-(\ref{phiso1}) 
with $\beta_4=0$ at $r=1.001r_h$. 
The quantities $-\alpha_6, \alpha_4, c_{r2}^2, c_{\Omega 2}^2$ are all positive throughout the horizon exterior. 
Note that $c_{r1}^2$ and $c_{\Omega 1}^2$ are 
equivalent to 1 for arbitrary $r$.
}
\end{figure}

In order to confirm the odd-parity stability of BHs 
in the intermediate regime between $r \simeq r_h$ and 
$r \gg r_h$, we numerically compute the quantities 
$-\alpha_6, \alpha_4$ and 
$c_{r2}^2, c_{\Omega2}^2$ outside 
the horizon by using Eqs.~(\ref{fho2})-(\ref{phiso1})
as boundary conditions around $r=r_h$. 
The numerical simulation of Fig.~\ref{fig1} corresponds to 
the coupling $\tilde{\beta}_3=1$ with $\mu=0.5$.
As we see in the left panel, both 
$-\alpha_6$ and $\alpha_4$ remain positive throughout  
the horizon exterior, so 
the no-ghost conditions are satisfied in this case.  

In the right panel of Fig.~\ref{fig1}, we observe that 
the deviation of the radial propagation speed squared 
$c_{r2}^2$ from 1 approaches 0 in the limit $r \to r_h$, 
while, at large distances ($r \gg r_h$), 
it decreases according to $c_{r2}^2-1=20\beta_3^2Q^2/r^6$. 
The angular propagation speed squared $c_{\Omega 2}^2$ 
exhibits the deviation from 1 at the horizon, such that 
$c_{\Omega 2}^2-1=4\mu (1-\mu) \tilde{\beta}_3^2
/[1+4\mu (1-\mu) \tilde{\beta}_3^2]>0$. 
Since $c_{\Omega 2}^2 -1 \simeq -4\beta_3^2 Q^2/r^6<0$
at spatial infinity, $c_{\Omega 2}^2$ crosses the value 1 
at an intermediate distance ($r \simeq 1.6r_h$ in Fig.~\ref{fig1}). Numerically we confirmed that both 
$c_{r2}^2-1$ and $|c_{\Omega 2}^2-1|$ 
are smaller than order 1 for 
$|\tilde{\beta}_3| \lesssim 1$, so 
there is no Laplacian instabilities of odd-parity perturbations 
outside the horizon.

Taking the limit $|\tilde{\beta}_3| \gg 1$ in Eqs.~(\ref{cr2es}) and (\ref{cO2es}), the asymptotic 
values of $c_{r2}^2$ and $c_{\Omega 2}^2$ 
at $r=r_h$ are 1 and 2, respectively. 
For $|\tilde{\beta}_3| \gtrsim 10$, our numerical simulations 
show that there are regions in which 
$c_{r2}^2, c_{\Omega 2}^2$ as well as $\alpha_4$ 
temporally become negative outside the horizon.
In such cases, the hairy BHs are unstable against odd-parity 
perturbations. In summary, as long as the cubic coupling 
is in the range 
\be
|\tilde{\beta}_3| \lesssim {\cal O}(1)\,,
\ee
there are neither ghost nor Laplacian instabilities throughout the horizon exterior. 

It is worthwhile to mention that the cubic interaction $f_3$ is 
related to the $g_5$ coupling of generalized Proca 
theories \cite{Heisenberg,Tasinato,Allys,Jimenez}.   
The Lagrangian $\mathcal{L}^3_{\rm SVT}$ 
with $\tilde{f}_3=0$ coincides with that of the $g_5$ 
coupling by replacing $\nabla_{\mu}\phi$ with $A_{\mu}$, 
in which case the scalar derivative $\phi'(r)$ is 
the placeholder of the longitudinal component of $A_{\mu}$. 

\subsection{$\beta_4 \neq 0$ and $n=0$}
\label{caseB}

Let us consider the theories in which $f_4$ is 
a nonvanishing constant $\beta_4$. 
{}From Eq.~(\ref{cr1s}), the radial propagation speed 
squared arising from the gravity sector yields
\be
c_{r1}^2=1\,.
\ee
On the other hand, the angular propagation speed squared 
(\ref{co1s}) reduces to 
$c_{\Omega 1}^2=1-4\beta_4h A_0'^2/(M_{\rm pl}^2 f)$, 
which is different from 1 unlike the theories 
with $\beta_4=0$. Moreover, the coupling $\beta_4$ 
gives rise to the value of $\alpha_6$ different from 
$-hM_{\rm pl}^2/(4r^4)$.

To estimate the quantities 
$\alpha_6, \alpha_4, c_{r2}^2, c_{\Omega 1}^2, 
c_{\Omega 2}^2$, we use the iterative solutions 
(\ref{fho2})-(\ref{phiso1}) in the vicinity of the horizon. 
Then, it follows that 
\ba
&&
\alpha_6 = -\frac{(1-\mu)\Mpl^2}{4r_h^4}
\left(1-\frac{8\tb_4\mu}{1+8\tb_4}\right)\left(\frac{r}{r_h}-1\right)
+{\cal O}((r/r_h-1)^{2})\,,\notag\\
&&
\alpha_4=\frac{1}{2r_h^2(1-\mu)}\left[1
-4(1-\mu) \left( \tb_4-\frac{\tb_3^2 \mu}{1+8\tb_4} \right)\right]\left(\frac{r}{r_h}-1\right)^{-1}
+{\cal O}((r/r_h-1)^{0})\,,\notag\\
&&
c_{r2}^2=1+{\cal O}(r/r_h-1)\,,\qquad 
c_{\Omega1}^2=1-\frac{8\tb_4\mu}{1+8\tb_4}+{\cal O}(r/r_h-1)\,,\notag\\
&&
c_{\Omega2}^2=
1+\frac{4\tb_3^2(1+8\tb_4+16\tb_4\mu)\mu(1-\mu)
+4\tb_4 (1+8\tb_4)[3-5\mu+8\tb_4(1-\mu)(3-2\mu)]}
{(1+8\tb_4)\left[4\tb_3^2\mu(1-\mu)+(1+8\tb_4)(1-4\tb_4+4\tb_4\mu)\right]}
+{\cal O}(r/r_h-1)\,.
\label{stabilityho}
\ea
For the odd-parity stability of BHs around 
the horizon, we require the following three conditions: 
\ba
& &
\left( 1+8\tb_4 \right) \left[1+8\tb_4(1-\mu) 
\right]>0 \,,\label{con1}\\
& &
\left( 1+8\tb_4 \right) \left[ 4\tb_3^2 \mu (1-\mu) 
+\left(1+8\tb_4 \right) \left\{ 1-4\tb_4 (1-\mu) \right\}
\right]>0\,, \label{con2}\\
& &
8\mu \left(1-\mu \right) \left[ 1+8\tb_4 (1+\mu) \right] \tb_3^2
+\left(1+8\tb_4 \right) \left[ 1+8\tb_4  (1-\mu)\right]^2 \geq 0
\,.\label{con3}
\ea
In the limit that $\tb_3 \to 0$, these conditions hold 
for $-1/8<\tb_4<1/[4(1-\mu)]$.  
This matches with the condition (6.8) of Ref.~\cite{KMTZ} 
derived for the sixth-order coupling 
$G_6=\beta_6={\rm constant}$ of 
$U(1)$ gauge-invariant generalized Proca 
theories with the branch $A_1=0$ 
after replacing $\tb_4$ with $\beta_6/(4r_h^2)$.
In this limit, there is only a vector hair associated 
with the temporal component $A_0$ as in the case 
of BH solutions advocated in Ref.~\cite{HorndeskiBH}.
The cubic coupling $\tb_3$ gives rise to the 
scalar hair with a nonvanishing value of $\phi'$. 
For $\tb_4>0$, the condition (\ref{con2}) gives the 
following bound 
\be
\tb_4<\frac{1+\mu+
\sqrt{(3-\mu)^2+32\tb_3^2 \mu (1-\mu)^2}}
{16(1-\mu)}\,,
\label{tb4}
\ee
whereas the other conditions (\ref{con1}) and 
(\ref{con3}) are automatically satisfied.  
The presence of cubic coupling $\tb_3$ leads to 
the larger upper limit of $\tb_4$ relative to 
the case $\tb_3=0$.

In the regime $r \gg r_h$, the leading-order terms of $\alpha_4$ and 
$c_{r2}^2$ are of the same forms as those given in 
Eqs.~(\ref{al4}) and (\ref{cr2d}), respectively. 
The quantities $\alpha_6, c_{\Omega 1}^2, c_{\Omega 2}^2$ have the following asymptotic behavior:
\ba
\alpha_6 
&=& -\frac{M_{\rm pl}^2}{4r^4}
+\frac{M_{\rm pl}^2 M}{2r^5}+{\cal O} 
\left( \frac{1}{r^6} \right)\,,\label{al6n}\\
c_{\Omega 1}^2 
&=& 1-\frac{4\beta_4 Q^2}{M_{\rm pl}^2 r^4}+{\cal O} 
\left( \frac{1}{r^5} \right)\,,
\label{cW1a}\\\
c_{\Omega 2}^2 
&=& 1+\frac{24\beta_4 M}{r^3}+{\cal O} 
\left( \frac{1}{r^4} \right)\,,
\label{cW2a}\
\ea
which show that the stability conditions against odd-parity perturbations are satisfied at spatial infinity.

\begin{figure}[h]
\begin{center}
\includegraphics[height=3.3in,width=3.3in]{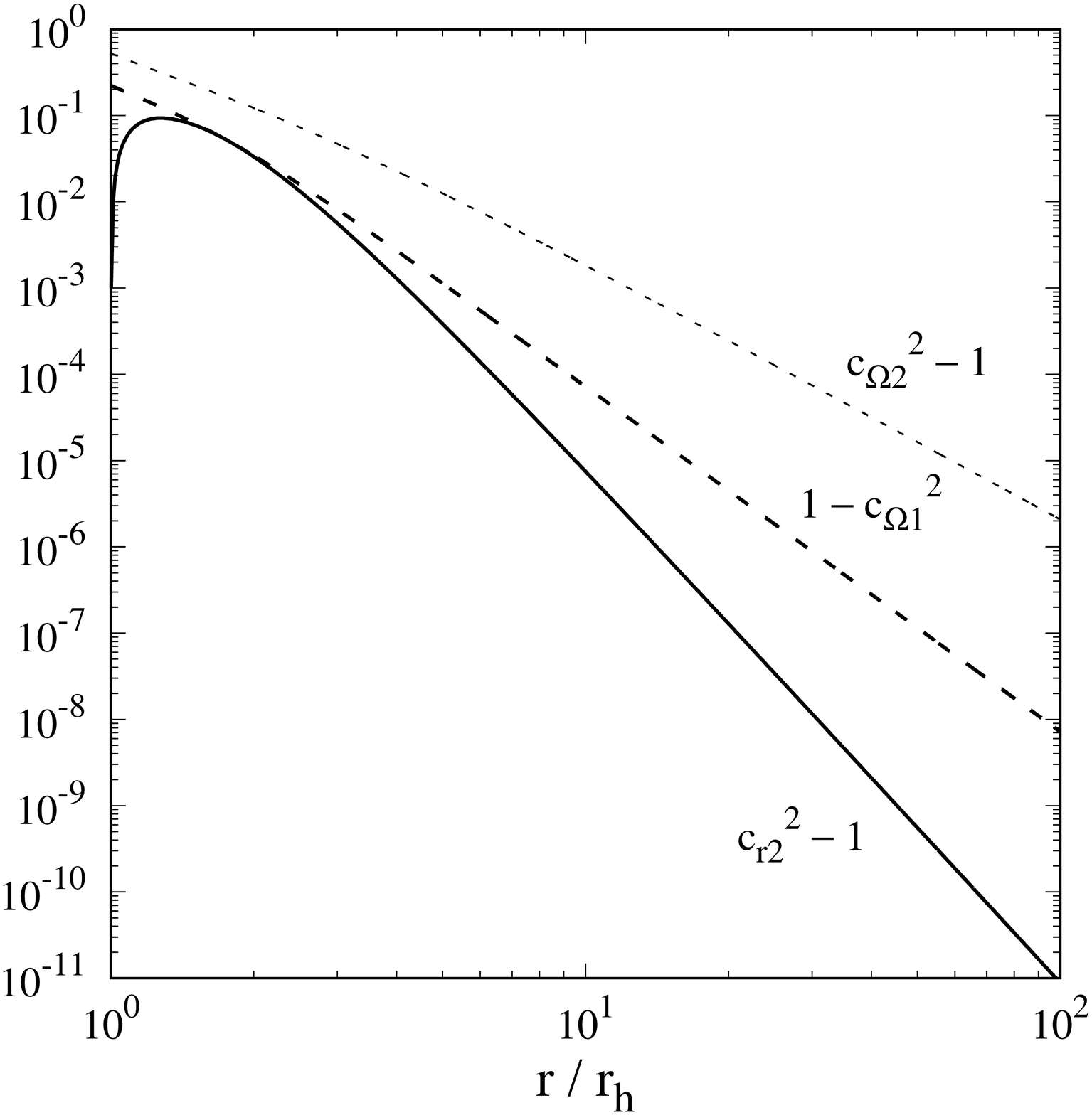}
\end{center}
\caption{\label{fig2}
Numerical solutions to $c_{r2}^2-1$, $1-c_{\Omega1}^2$, 
$c_{\Omega2}^2-1$ versus $r/r_h$ for the model (\ref{conmodel}) 
with $\tb_3=0.5$, $\tb_4=0.1$, $n=0$, and  $\mu=0.5$.
}
\end{figure}

In Fig.~\ref{fig2}, we plot the deviations of 
$c_{r2}^2,c_{\Omega1}^2,c_{\Omega2}^2$ from 1 as functions of 
$r/r_h$ for $\tb_3=0.5$, $\tb_4=0.1$ 
and $\mu=0.5$. These model parameters are chosen to be 
consistent with the conditions (\ref{con1})-(\ref{con3}).
As estimated from Eq.~(\ref{stabilityho}), the numerical simulation 
of Fig.~\ref{fig2} shows that the angular propagation 
speed squares on the horizon are in the ranges $c_{\Omega1}^2<1$ and $c_{\Omega2}^2>1$, while $c_{r2}^2 \to 1$ as $r\to r_h$. 
For the distance $r \gg r_h$, the deviations of propagation speed 
squares from 1 rapidly decrease as $c_{r2}^2-1\propto r^{-6}$, 
$1-c_{\Omega1}^2\propto r^{-4}$, and $c_{\Omega2}^2-1\propto r^{-3}$, 
whose properties agree with the analytic estimations given in Eqs.~(\ref{cr2d}), (\ref{cW1a}) and (\ref{cW2a}). 
For the model parameters chosen in Fig.~\ref{fig2}, 
there are no Laplacian instabilities outside the horizon.
We also confirmed that the quantities 
$-\alpha_6$ and $\alpha_4$ are positive throughout the horizon 
exterior, so the conditions for the absence of ghosts are satisfied. 

Provided that the positive coupling $\tb_4$ is within the 
range (\ref{tb4}), the ghosts and 
Laplacian instabilities do not typically arise for 
$|\tb_3| \lesssim {\cal O}(1)$. 
For $|\tb_3|\gtrsim10$, the quantities 
$c_{r2}^2, c_{\Omega2}^2$ as well as $\alpha_4$ temporally 
become negative in the region not far from the horizon.
This property is similar to the case (A) discussed 
in Sec.~\ref{caseA}.

\subsection{$\beta_4 \neq 0$ and $n=1$}

Let us finally proceed to the quartic coupling $f_4(X)=\beta_4 X$.
Unlike the model (B), the model (C) contains a nonminimal 
coupling with an explicit interaction with the scalar 
derivative $\phi'$.
In this case, the propagation speed squares $c_{r1}^2$ and $c_{\Omega 1}^2$ are different from 1.

In the vicinity of the horizon, we use the iterative solutions 
(4.15)-(4.18) of Ref.~\cite{HT18} to compute the quantities 
$\alpha_6, \alpha_4, c_{r1}^2, c_{r2}^2, c_{\Omega 1}^2, 
c_{\Omega 2}^2$. Then, it follows that 
\ba
& &
\alpha_6 =-\frac{M_{\rm pl}^2 (1-\mu)}{4r_h^4} 
\left( \frac{r}{r_h}-1 \right)+{\cal O}((r/r_h-1)^{2})\,,
\nonumber \\
& &
\alpha_4= \frac{1+4\tb_3^2 \mu(1-\mu)+8\bar{\beta}_4 \mu}
{2r_h^2(1-\mu)(1+8\bar{\beta}_4 \mu)}\left(\frac{r}{r_h}-1\right)^{-1}
+{\cal O}((r/r_h-1)^{0})\,, 
\nonumber \\
& &
c_{r1}^2=1+{\cal O}(r/r_h-1)\,,\qquad 
c_{r2}^2=1+{\cal O}(r/r_h-1)\,,\qquad
c_{\Omega1}^2=1+{\cal O}(r/r_h-1)\,,\nonumber \\
& &
c_{\Omega2}^2=
1+\frac{8\tb_3^2\mu(1-\mu)[1+2\bb_4\mu(5-\mu)]}
{2(1+8\bb_4\mu)[1+4\tb_3^2\mu(1-\mu)+8\bb_4\mu]}
+{\cal O}(r/r_h-1)\,, 
\label{stabilityho2}
\ea
where $\bar{\beta}_4 \equiv \beta_4 M_{\rm pl}^2/r_h^4$. 
Then, the conditions 
$c_{r1}^2 \geq 0$, $c_{r2}^2 \geq 0$, and 
$c_{\Omega 1}^2 \geq 0$ hold around $r=r_h$. 
{}From the requirements $\alpha_4>0$ and 
$c_{\Omega 2}^2 \geq 0$, we obtain the 
following bounds:
\ba
& &
[1+4\tb_3^2 \mu(1-\mu)+8\bar{\beta}_4 \mu]
(1+8\bar{\beta}_4 \mu) > 0\,,\label{cond1}\\
& &
1+8\tb_3^2 \mu (1-\mu)
[1+\bar{\beta}_4\mu (9-\mu)]+16\bar{\beta}_4 \mu
(1+4\bar{\beta}_4 \mu) \geq 0\,.\label{cond2}
\ea
In the limit that $\bar{\beta}_4 \to 0$, these conditions 
are trivially satisfied. This property also persists 
in another limit $\tb_3 \to 0$, in which case the two terms on 
the left hand sides of Eqs.~(\ref{cond1}) and (\ref{cond2}) 
reduce to $(1+8\bar{\beta}_4 \mu)^2$ with 
$c_{\Omega2}^2 \to 1+{\cal O}(r/r_h-1)$.
In contrast to the case (B) discussed in Sec.~\ref{caseB}, 
the conditions (\ref{cond1}) and (\ref{cond2}) automatically 
hold for positive $\bb_4$ and hence there is no upper 
bound of $\bb_4$.

On using the iterative solutions (3.19)-(3.22) of 
Ref.~\cite{HT18} in the regime $r \gg r_h$, 
we find that the leading-order 
terms of $\alpha_4, c_{r2}^2, c_{\Omega 2}^2, \alpha_6$ are 
of the same forms as Eqs.~(\ref{al4}), (\ref{cr2d}), (\ref{cW2}), and (\ref{al6n}) respectively. 
The propagation speed squares $c_{r1}^2$ and 
$c_{\Omega 1}^2$ have the following asymptotic behavior:
\be
c_{r1}^2=1+\frac{4\beta_3^2 \beta_4 Q^6}
{M_{\rm pl}^2r^{14}}+
{\cal O}\left( \frac{1}{r^{15}} \right)\,,\qquad 
c_{\Omega 1}^2=1+\frac{12\beta_3^2 \beta_4 Q^6}
{M_{\rm pl}^2r^{14}}+
{\cal O}\left( \frac{1}{r^{15}} \right)\,,
\label{c1inf}
\ee
so there are neither ghost nor Laplacian 
instabilities at spatial infinity.

\begin{figure}[h]
\begin{center}
\includegraphics[height=3.3in,width=3.3in]{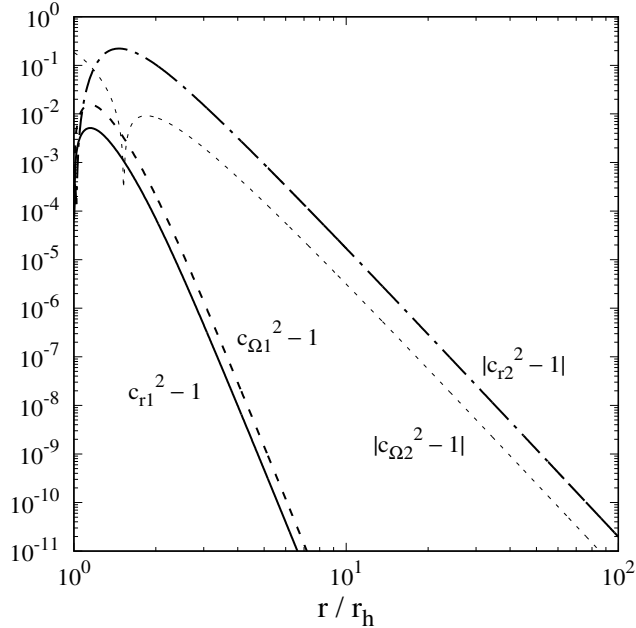}
\end{center}
\caption{\label{fig3}
Numerical solutions to $c_{r1}^2-1$, $|c_{r2}^2-1|$, 
$c_{\Omega1}^2-1$, $|c_{\Omega2}^2-1|$ versus $r/r_h$
for the model (\ref{conmodel}) with $\tb_3=1$, 
$\bar{\beta}_4=1$, $n=1$, and $\mu=0.5$.
}
\end{figure}

In Fig.~\ref{fig3}, we show numerical solutions to the deviations 
of propagation speed squares $c_{r1}^2$, $c_{r2}^2$, 
$c_{\Omega1}^2$, $c_{\Omega2}^2$ from 1 for the model parameters 
$\tb_3=1$, $\bar{\beta}_4=1$, and $\mu=0.5$. 
The asymptotic behavior of those quantities 
around $r=r_h$ and $r \gg r_h$ is consistent with the analytic 
estimations given in Eqs.~(\ref{stabilityho2}) and (\ref{c1inf}).
Since $c_{r1}^2-1$, $|c_{r2}^2-1|$, $c_{\Omega1}^2-1$, 
$|c_{\Omega2}^2-1|$ remain smaller than 1, 
Laplacian instabilities are absent throughout the 
horizon exterior. We also numerically confirmed that 
no-ghost conditions hold for the model parameters 
used in Fig.~\ref{fig3}.

If the quartic coupling is in the range $\bar{\beta}_4 \lesssim 
{\cal O}(1)$, all the stability conditions can be 
consistently satisfied for $|\tb_3|\lesssim{\cal O}(1)$, but for 
$|\tb_3| \gtrsim 10$ there are intermediate regions outside 
the horizon in which $\alpha_4,c_{r2}^2,c_{\Omega2}^2$ become 
negative. This situation is analogous to what we discussed 
in Secs.~\ref{caseA} and \ref{caseB}. 
Unlike the case (B), however, the coupling $\beta_4$ is not 
bounded from above. If the couplings range in the region 
$\bar{\beta}_4 \gg \tb_3^2$, we find that the unstable 
regions tend to disappear even for $|\tb_3| \gtrsim 10$.
In this case, the cubic coupling $|\tb_3|$ is effectively 
negligible relative to $\bar{\beta}_4$ in 
Eqs.~(\ref{stabilityho2}) and (\ref{c1inf}), 
so all the propagation speed squares are close to 1 in 
two asymptotic regimes. Indeed, this is also the case 
at intermediate distances with no-ghost conditions satisfied. 
In summary, the model (C) allows 
the possibility for satisfying stability conditions
for wider ranges of $\beta_3$ and $\beta_4$ than those 
in the cases (A) and (B).

\section{Conclusions}
\label{concludesec}

The SVT theories correspond to the unified framework of most general 
scalar-tensor (Horndeski) and vector-tensor (generalized Proca) theories 
with second-order equations of motion. If the theories respect the 
$U(1)$ gauge symmetry, the new interactions arising in such theories 
are given by the Lagrangians (\ref{L2})-(\ref{L4}). 
Because of the gauge invariance, there is no longitudinal propagation 
of the vector field $A_{\mu}$. Hence the $U(1)$ gauge-invariant 
SVT theories contain the five dynamical DOFs: one scalar, two 
transverse vector modes, and two tensor polarizations.

If we apply the $U(1)$ gauge-invariant SVT theories to vacuum solutions 
on the static and spherically symmetric background, 
the cubic-order coupling $f_3=\beta_3={\rm constant}$ 
can give rise to hairy BH solutions. 
For the couplings (\ref{f24}), the iterative BH 
solutions around the horizon are given by 
Eqs.~(\ref{fho2})-(\ref{phiso1}), whereas 
the solutions at spatial infinity are of the forms
(\ref{fho2d})-(\ref{phiso2}). The coupling $\beta_3$ 
generates a scalar hair with a nonvanishing field derivative 
$\phi'$. Besides this cubic-order interaction, the quartic coupling $\beta_4$ 
also leads to modifications to the RN solution through a vector hair.
The effect of couplings $\beta_3$  and $\beta_4$ on the metric 
components $f$ and $h$ mostly 
manifests themselves in the vicinity of the horizon. 

In this paper, we provided a general framework for 
studying the stability of static and spherically symmetric BHs 
against odd-parity perturbations in $U(1)$ gauge-invariant SVT theories. 
For the modes $l \geq 2$, the dynamical fields correspond to 
the perturbation $\chi$ defined by Eq.~(\ref{chi}) arising from the 
gravity sector and the vector-field perturbation $\delta A_{lm}$. 
We showed that, under the conditions $\alpha_6<0$ and $\alpha_4>0$,  these perturbations do not contain ghost modes.
The radial propagation speed squares $c_{r1}^2$ and 
$c_{r2}^2$ associated with the perturbations
$\chi$ and $\delta A_{lm}$ are given, respectively, by 
Eqs.~(\ref{cr1}) and (\ref{cr2}).
In the limit that $L=l(l+1) \gg 1$, we also derived the 
angular propagation speed squares $c_{\Omega 1}^2$ 
and $c_{\Omega 2}^2$ in the forms (\ref{co1}) and (\ref{co2}), 
respectively. As we observe in Eqs.~(\ref{cr1s}) and (\ref{co1s}),
the quartic couplings $f_4$ and $\tilde{f}_4$ can 
lead to the deviations of $c_{r1}^2$ and $c_{\Omega 1}^2$ 
from 1. For the dipole mode $l=1$, there is only 
the vector-field perturbation $\delta A_{lm}$, whose stability 
conditions are the same as those for $l \geq 2$.

We applied stability conditions of odd-parity perturbations 
to concrete models given by the couplings (\ref{conmodel}). 
In the absence of quartic couplings ($\beta_4=0$), 
the propagation speed squares $c_{r1}^2$ and $c_{\Omega 1}^2$ 
are equivalent to 1 with $\alpha_6=-hM_{\rm pl}^2/(4r^4)$, 
so there are neither ghost nor Laplacian instabilities outside 
the horizon ($h>0$) for the perturbation in the gravity sector. 
In this case, the stability conditions 
$\alpha_4>0, c_{r2}^2 \geq 0, c_{\Omega 2}^2 \geq 0$ 
are satisfied both in the near-horizon limit and 
at spatial infinity. However, for $|\tilde{\beta}_3| \gtrsim 10$, 
we find that these conditions can be violated in an intermediate regime 
between $r=r_h$ and $r \gg r_h$. Hence the cubic coupling should be 
in the range $|\tilde{\beta}_3| \lesssim {\cal O}(1)$ to ensure the 
odd-parity stability of vector-field perturbations. 

The quartic couplings generally modify stability conditions 
of odd-parity perturbations relative to the case $\beta_4=0$, 
but we showed that there are viable model parameter spaces 
in which there are neither ghost nor Laplacian instabilities 
throughout the horizon exterior.
For the quartic interaction $f_4=\beta_4={\rm constant}$, 
the coupling $\beta_4$ consistent with stability conditions 
is bounded from above, 
but this is not the case for $f_4=\beta_4 X$. 
Hence the latter model leads to a wider allowed range of 
couplings $\beta_3$ and $\beta_4$ relative to the former model.

In this paper we studied the BH stability against 
odd-parity perturbations, but it is of interest to extend 
our analysis to even-parity perturbations.
There are one scalar, one vector, and one tensor modes 
arising from the even-parity sector.
In particular, it remains to be seen whether the existence 
of scalar perturbations puts further constraints on hairy 
BH solutions present in SVT theories. 
Moreover, it will be interesting to investigate
the speeds of tensor and vector perturbations arising from 
the even-parity sector. One could also put further constraints for the absence of superluminal propagation since one might worry about the construction of closed time-like curves and acausality
(even though a similar chronology protection might occur 
as in Galileon theories \cite{Burrage:2011cr}).
These issues are left for future works.

\section*{Acknowledgements}

We are grateful to Masashi Kimura and 
Masato Minamitusji for fruitful discussions.
LH thanks financial support from Dr.~Max R\"ossler, 
the Walter Haefner Foundation and the ETH Zurich
Foundation.  
RK is supported by the Grant-in-Aid for Young 
Scientists B of the JSPS No.\,17K14297. 
ST is supported by the Grant-in-Aid 
for Scientific Research 
Fund of the JSPS No.~16K05359 and 
MEXT KAKENHI Grant-in-Aid for 
Scientific Research on Innovative Areas ``Cosmic Acceleration'' (No.\,15H05890). 
ST thanks warm hospitalities to CENTRA and 
ETH-ITS Zurich in which a part of this work was done.


\end{document}